\documentclass{emulateapj}
\usepackage{apjfonts}

\usepackage{psfig}

\journalinfo{Accepted by the Astronomical Journal}

\shorttitle{A Search for Rapidly Spinning Pulsars and Fast Transients}

\shortauthors{Schmidt et al.}

\begin{document}

\title{A Search for Rapidly Spinning Pulsars and Fast Transients in
Unidentified Radio Sources with the NRAO 43-Meter Telescope}

\author{Deborah Schmidt\altaffilmark{1,2,3}, Fronefield
Crawford\altaffilmark{1}, Glen Langston\altaffilmark{2}, \& Claire
Gilpin\altaffilmark{1},}

\altaffiltext{1}{Department of Physics and Astronomy, Franklin and
Marshall College, P.O. Box 3003, Lancaster, PA 17604, USA}

\altaffiltext{2}{National Radio Astronomy Observatory, P.O. Box 2,
Green Bank WV 24944}

\altaffiltext{3}{Department of Astronomy, University of Arizona, 933
North Cherry Avenue, Tucson, AZ 85721, USA}


\begin{abstract}
We have searched 75 unidentified radio sources selected from the NRAO
VLA Sky Survey (NVSS) catalog for the presence of rapidly spinning
pulsars and short, dispersed radio bursts.  The sources are radio
bright, have no identifications or optical source coincidences, are
more than 5\% linearly polarized, and are spatially unresolved in the
catalog.  If these sources are fast-spinning pulsars
(e.g. sub-millisecond pulsars), previous large-scale pulsar surveys
may have missed detection due to instrumental and computational
limitations, eclipsing effects, or diffractive scintillation. The
discovery of a sub-millisecond pulsar would significantly constrain
the neutron star equation of state and would have implications for
models predicting a rapid slowdown of highly recycled X-ray pulsars to
millisecond periods from, e.g., accretion disk decoupling.  These same
sources were previously searched unsuccessfully for pulsations at 610
MHz with the Lovell Telescope at Jodrell Bank. This new search was
conducted at a different epoch with a new 800 MHz backend on the NRAO
43-meter Telescope at a center frequency of 1200 MHz. Our search was
sensitive to sub-millisecond pulsars in highly accelerated binary
systems and to short transient pulses. No periodic or transient
signals were detected from any of the target sources. We conclude that
diffractive scintillation, dispersive smearing, and binary
acceleration are unlikely to have prevented detection of the large
majority of the sources if they are pulsars, though we cannot rule out
eclipsing, nulling or intermittent emission, or radio interference as
possible factors for some non-detections. Other (speculative)
possibilities for what these sources might be include radio-emitting
magnetic cataclysmic variables or older pulsars with aligned magnetic
and spin axes.
\end{abstract}


\section{Introduction}

Radio pulsars have a wide range of spin periods, but no pulsar has yet
been observed to be rotating faster than PSR J1748$-$2446ad, which has
a spin period of $P = 1.39$ ms \citep{hrs+06}.  Models of the neutron
star equation of state (EOS) and the properties of pulsars that have
been spun up (recycled) via accretion from a companion suggest that
pulsars with spin periods less than a millisecond (``sub-millisecond''
pulsars) can exist structurally and could in principle be produced
from the recycling process (e.g., Cooke et al. 1994\nocite{cst94};
Haensel et al. 1999\nocite{hlz99}). A significant population of
sub-millisecond pulsars may therefore be present in the Galaxy waiting
to be discovered (see, e.g., Possenti et
al. 1998\nocite{pcd+98}). However, some studies have suggested that a
spin equilibrium is reached during accretion which limits the neutron
star spin to larger (millisecond) periods. In these scenarios, the
spin up from accretion is balanced by angular momentum loss through,
e.g., an accretion disk/magnetosphere interaction \citep{phd12} or the
emission of gravitational radiation \citep{w84,b98}. \citet{hma11}
provide a brief overview.  A more recent model presented by
\citet{t12} suggests that highly recycled neutron stars can be
produced as sub-millisecond X-ray sources during the accretion phase,
but that they are then rapidly spun down in the Roche lobe decoupling
phase. In this case, the magnetosphere expands during the last phase
of accretion, a large fraction of the rotational energy is dissipated,
and the pulsar is not seen as a sub-millisecond pulsar when the radio
emission turns on.

The discovery of a sub-millisecond pulsar would be of particular
importance since it would falsify models suggesting that
sub-millisecond radio pulsars should not be observed (see above) while
also placing significant constraints on the largely unknown EOS of
neutron star matter (e.g., Lattimer \& Prakash
2004\nocite{lp04}). However, sub-millisecond pulsars are difficult to
detect owing to a number of selection effects. These include
limitations in observing sensitivity, digital sampling rates, and
frequency channel resolution, as well as eclipsing and acceleration
effects if the pulsar is in a tight binary system. In fact, very
rapidly spinning pulsars may be preferentially located in such
eclipsing systems \citep{hrs+06,h08}, making such systems even harder
to detect.  Large-area surveys having the necessary time and frequency
resolution (as well as the required post-observation computational
power for data processing) to detect such pulsars have not yet been
feasible on the full sky. Therefore, targeting smaller areas likely to
harbor pulsars remains one of the most promising avenues for trying to
discover a sub-millisecond pulsar. Examples of this approach include
the highly successful deep targeted radio searches of {\it Fermi}
$\gamma$-ray sources \citep{rap+12,rrc+13} in which an astonishing 44
new millisecond pulsars (MSPs) have recently been discovered.

Some pulsars are also known to exhibit detectable single pulses at
irregular intervals. Notable examples of this type include the Crab
pulsar \citep{sr68} and the MSP PSR B1937+21 \citep{cst+96}, both of
which are also detectable as periodic sources. Other, more recently
discovered sources of sporadic radio emission include rotating radio
transients (RRATs; McLaughlin et al. 2006\nocite{mll+06}), which are
rotating neutron stars that are only detectable as transient burst
sources. 

In an effort to discover previously undetected millisecond and
sub-millisecond pulsars, as well as to find new transient emitters, we
have searched a sample of unidentified radio sources from the NRAO VLA
Sky Survey (NVSS) \citep{ccg+98}. These sources were originally
selected by \citet{ckb00} for a similar pulsar search with the 76-m
Lovell Telescope at the Jodrell Bank Observatory. We outline the
source selection criteria below, which can also be found in
\citet{ckb00}. The NVSS is a large-scale radio survey of the northern
sky at 1400 MHz conducted with the Very Large Array (VLA). It covered
declinations $\delta > -40^{\circ}$ in D and DnC
configurations with a synthesized beam size of 45$''$. Position errors
in the NVSS catalog are typically $\la 1''$ for strong sources. The
catalog contains $\sim 3 \times 10^{5}$ sources with a flux density
greater than 15 mJy, when local sources and unreliable survey regions
are excluded (see, e.g., Blake \& Wall 2002\nocite{bw02}; Crawford
2009\nocite{c09}). Many of the sources detected in the NVSS remain
unidentified.

Owing to the large number of extended, resolved sources in the
catalog, it was necessary to explicitly check for the point-like
nature of those objects. Sources were checked for corresponding
detections in the Faint Images of the Radio Sky at Twenty Centimeters
(FIRST) survey \citep{bwh95}. The FIRST survey covered the north and
south Galactic caps using the VLA in B configuration with a
synthesized beam size of 5$''$.4.  For sources found in both the NVSS
and FIRST catalogs, the corresponding flux densities were compared. If
a source was extended, the better resolution of the FIRST survey would
result in a lower flux density than the NVSS survey, since some of the
flux would be resolved out. Therefore, sources were only included if
their FIRST and NVSS flux densities agreed to within a few percent
(indicating an unresolved, non-variable source) or if the FIRST flux
density exceeded the NVSS flux density (indicating an unresolved
scintillating source). For those sources outside of the FIRST survey
region, observations were performed using the VLA in B configuration
to achieve the same angular resolution as the FIRST survey. The flux
densities of the sources observed in these pointings were then
compared with their NVSS values to eliminate any additional extended
sources.

The target list was further reduced by requiring that the sources have
no identifications with known sources (and no optical source
coincidences), be radio bright (with a 1400 MHz flux density $\ge$ 15
mJy), be spatially unresolved in the NVSS catalog, and be more than
5\% linearly polarized. The polarization criterion was used since
pulsars often have a high degree of linear polarization \citep{lm88},
which makes polarized point radio sources good candidates to
search. Although there is not a clear polarization cutoff separating
the pulsar and extragalactic populations, a polarization threshold of
5\% excludes about 90\% of the identified non-pulsar population while
retaining about 90\% of the identified pulsar population \citep{ht99,
ckb00}.

Using these criteria, a list of 92 target sources was compiled and
searched by \citet{ckb00} for radio pulsations using the 76-m Lovell
Telescope. No pulsations were detected from any of the target sources
in that search, but no search for single pulses was done at that
time. A recent re-analysis of these data using a search for single
pulses revealed no new detections. Since the completion of the
\citet{ckb00} survey, 16 of the 92 target sources have been identified
as extragalactic objects in
SIMBAD\footnote{http://simbad.u-strabg.fr/simbad} (see Table
\ref{tbl-1}). Some of the targets have also been detected in other
radio surveys (listed in Table \ref{tbl-1}) at different
frequencies. In these cases we measured spectral indices for them (see
Table \ref{tbl-1}). All of the measured spectral index values are
shallower than the average pulsar spectral index of $\alpha = -1.6$
\citep{lyl+95}, where $\alpha$ is defined according to $S \sim
\nu^{\alpha}$ ($S$ is flux density and $\nu$ is the observing
frequency). This does not preclude these sources being pulsars, but
they are not as steep as expected.

For our search, we have used the same set of selection criteria but
have excluded the 16 sources that now have secure extragalactic
identifications. One remaining source was not observed owing to
telescope scheduling limitations, leaving 75 unidentified targets that
we observed and searched.  We checked the {\it Fermi} LAT 2-year Point
Source Catalog (2FGL) \citep{2fgl} for coincidences with our targets
and found that none lie within the 95\% error ellipses of any of the
{\it Fermi} sources.  Table \ref{tbl-1} presents the full list of 92
sources from Table 1 of \citet{ckb00}. The 16 secure extragalactic
identifications are indicated, as are the measured radio spectral
index values (where available). This list represents just a small
fraction ($< 0.1$\%) of the total number of sources in the NVSS
catalog with flux densities greater than 15 mJy.  Note that the flux
densities and polarization fractions listed were obtained from
revision 2.18 of the NVSS
catalog\footnote{http://www.cv.nrao.edu/nvss/NVSSlist.shtml} and in
some cases are slightly different from the earlier catalog values
presented in Table 1 of \citet{ckb00}.

\section{Motivation for New Observations}

\begin{figure}
\centerline{\psfig{figure=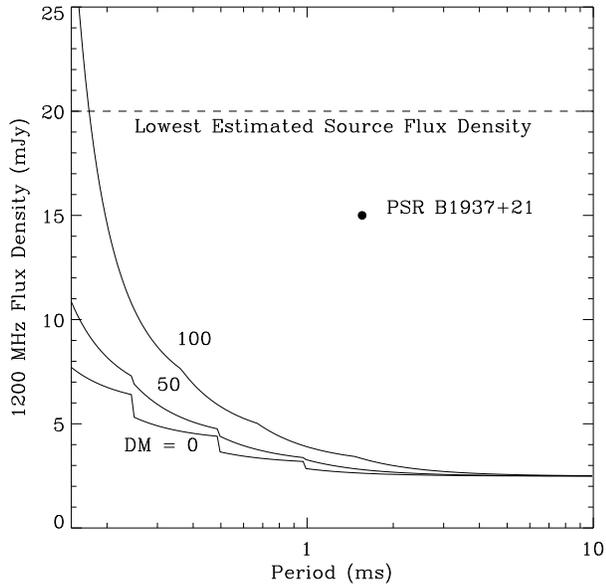,width=3.5in,angle=0}}
\caption{Sensitivity curves for our pulsar survey for assumed DMs of
0, 50, and 100 pc cm$^{-3}$ and a 5\% intrinsic pulsed duty
cycle. Each curve represents the flux density as a function of spin
period that would result in a S/N of 7 in the Fourier spectrum in the
search. The dashed line at 20 mJy indicates the lowest estimated flux
density of the 75 sources at 1200 MHz. The sensitivity baseline was
empirically determined by scaling a S/N = 33 blind detection of PSR
B1937+21, which has a DM of 71 pc cm$^{-3}$ and an estimated 1200 MHz
flux density of 15 mJy. Not included here is the additional selection
criterion that Fourier candidates must have $P$ $\ge$ 0.5 ms to be
considered, which effectively restricts our sensitivity to $P$ $\ge$
0.5 ms.\label{fig-1}}
\end{figure}

Starting in February of 2010, we conducted new observations of the 75
unidentified sources with the NRAO 43-meter
Telescope\footnote{http://www.gb.nrao.edu/43m/} at the Green Bank
Observatory. Several factors motivated this.  The first was the
availability of the
PRESTO\footnote{http://www.cv.nrao.edu/$\sim$sransom/presto} and
SIGPROC\footnote{http://sigproc.sourceforge.net} software packages,
which are now widely used for pulsar data analysis \citep{r01,rem02}.
PRESTO and the single pulse search capability of SIGPROC were not
available at the time of the search of \citet{ckb00}.  The development
of the Fourier-based search code in PRESTO for acceleration searches
was also a major advancement for detecting binary pulsar systems in
our search.  The ability to conduct acceleration searches of the
survey data is especially important for the detection of millisecond
and sub-millisecond pulsars since approximately 80\% of MSPs are known
to be in binaries (e.g., Lorimer 2008\nocite{l08}).  Second, in the
time that has passed since the original search, greatly increased
computational power for acceleration searches has become feasible at a
reasonable cost.  Finally, a new high bandwidth (800 MHz) backend with
fast sampling capability became available at the telescope (the West
Virginia Ultimate Pulsar Processing Instrument, or WUPPI; see, e.g.,
Mickaliger et al. 2012\nocite{mml+12}).  Sub-millisecond pulsars are
in principle detectable with this instrument, and the wide bandwidth
and many narrow frequency channels allow radio frequency interference
(RFI) to be more easily recognized and excised. The 8-bit sampling and
polyphase filterbank of WUPPI also mitigated RFI and reduced
digitization loss compared to the 1-bit sampling in the \citet{ckb00}
search.  The higher central observing frequency (1200 MHz versus 610
MHz for the previous survey) also helped reduce pulse broadening from
interstellar dispersion and scattering. Table \ref{tbl-2} summarizes
the different observing parameters for the two surveys.  For our
observing system, the dispersion broadening within channels was $\sim$
0.1 ms for DM = 100 pc cm$^{-3}$. This value is comparable to the
previous Jodrell Bank search observations which had narrower channels
but also a lower central observing frequency \citep{ckb00}. As seen in
Figure \ref{fig-1}, dispersion broadening does not significantly
reduce the sensitivity of our search even at very low periods ($P$
$\la$ 1 ms). Our survey maintains good sensitivity to sub-millisecond
pulsars for dispersion measures (DMs) between 0 and 100 pc cm$^{-3}$,
the range which is most relevant for our survey.

Apart from the effects of acceleration, the presence of orbital
companions can significantly affect detectability through eclipsing by
the companion.  Material blown off of a companion star by the pulsar
wind can obscure the pulsar signal in such cases \citep{t91,hrs+06}.
This has been observed for a number of MSPs with non-degenerate
companions, including PSR J1748-2446ad \citep{hrs+06} and PSR
J1740-5340 \citep{dpm+01}, both of which are located in globular
clusters and are eclipsed for $\sim$ 40\% of their orbit. The number
of ``black widow'' \citep{fst88} and ``redback'' \citep{asr+09}
eclipsing pulsar systems in the Galaxy has also recently grown (see
Roberts 2012\nocite{r12} for a summary). These pulsars are in tight
binary systems with eclipse fractions that can vary with frequency and
which can be quite large.  As mentioned above, eclipsing systems may
preferentially harbor very rapidly spinning pulsars \citep{h08},
making sub-millisecond pulsar systems even more likely to be missed.
Such systems might be discovered by re-observing them at a different
epoch when they are not in eclipse. Time-dependent effects such as
diffractive scintillation or sporadic radio emission from the neutron
star may also have previously prevented detection of some of our
sources. The much larger observing bandwidth in the search described
here helps mitigate diffractive scintillation effects (see Section 5
for more details).

For these reasons we re-performed a search of these targets using the
43-m Telescope and WUPPI backend. Details of our observational method
are provided in Section 3, followed by a description of our data
reduction techniques for both the periodicity search and the single
pulse search in Section 4. The results of our search are discussed in
Section 5, and our conclusions are presented in Section 6.

\section{Observations}

Observations of the 75 sources were conducted with the 43-m Telescope
and WUPPI backend between February 2010 and August 2011. We observed
each source at a center frequency of 1200 MHz in two orthogonal linear
polarizations for a total of 900 s per source. A bandwidth of 800 MHz
was split into 4096 frequency channels, and each channel was 8-bit
sampled every 61.44 $\mu$s.  Data were recorded directly to disk in
PSRFITS format \citep{hvm04}. We determined an empirical flux density
limit to pulsed emission for the search using a 900 s test observation
of the bright MSP PSR B1937+21. With a DM of 71 pc cm$^{-3}$ and a
period of 1.56 ms, PSR B1937+21 is similar to the rapidly spinning
pulsars we hoped to detect. According to the ATNF catalog
\citep{mht+05}\footnote{http://www.atnf.csiro.au/people/pulsar/psrcat},
PSR B1937+21 has a flux density of 10 mJy at 1400 MHz and a spectral
index of $\alpha$ = $-$2.6.  This corresponds to a flux density of 15
mJy at 1200 MHz, our central observing frequency.  The weakest of our
75 target sources had a flux density of 15 mJy at 1400 MHz (see Table
\ref{tbl-1}). Assuming a typical value of $\alpha$ = $-$1.6 for
pulsars \citep{lyl+95}, this scales to 20 mJy at 1200 MHz. A 900 s
observation of PSR B1937+21 yielded a clear detection of the pulsed
emission in the Fourier search with a signal-to-noise ratio (S/N) of
33 in the spectrum. Given the same observing configuration and
integration time, the S/N of our weakest source would be expected to
have an S/N of 44 in the absence of time-dependent variability in
intensity or Fourier bin drift due to acceleration (see Section
4.1). Assuming a S/N threshold of 7 for detectability, it is clear
that even our weakest source should be well above the limiting flux
density of our observations if it is a pulsar (see Figure
\ref{fig-1}).

Each set of observations in an observing session was preceded by an
approximately 90 s observation of a known bright pulsar in order to
ensure that the telescope and observing software were functioning
properly.

\section{Data Reduction and Analysis}

We searched each data set for both fast periodic signals and single
dispersed pulses at a range of DMs. The single pulse search was
employed specifically to detect transient objects that would not
appear in the periodicity search, and the periodicity search included
a search for accelerated signals to account for binary motion.  We
describe these searches below.

\subsection{Periodicity Search}
 
\begin{figure}
\centerline{\psfig{figure=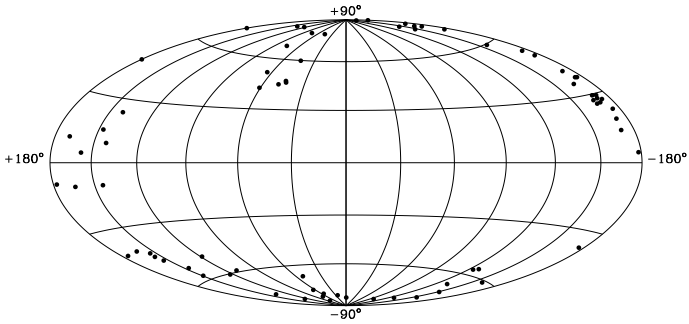,width=3.5in,angle=90}}
\caption{Aitoff projection plot of the sky positions of the 75
unidentified sources surveyed plotted in Galactic coordinates. The
Galactic plane runs horizontally and bisects the plot. Most of the
sources are located far from the Galactic plane, where there is less
ISM plasma, and subsequently the maximum expected DMs are
smaller. From the NE2001 model of \citet{cl02}, the estimated maximum
DM along the line of sight for all but 8 of the 75 sources is expected
to be less than 100 pc cm$^{-3}$ \citep{cl02}.\label{fig-2}}
\end{figure}

We performed the periodicity search using PRESTO.  Each observation
was first checked for bright narrowband and undispersed short-duration
signals, indicative of terrestrial RFI. Corrupted samples in the
time-frequency array were replaced by median value
powers. Additionally, we flagged several bands in the frequency range
in which RFI was known to be persistent. On average, we masked
approximately 10\% of the total data in each beam.  We dedispered the
data at 401 evenly spaced trial DMs between 0 and 100 pc cm$^{-3}$,
corresponding to a step size of 0.25 pc cm$^{-3}$. We chose this DM
range since the NE2001 model of the Galactic electron distribution
\citep{cl02} indicates that all but 8 of the 75 sources observed are
expected to have a maximum possible DM contribution along the line of
sight of less than 100 pc cm$^{-3}$, while the remaining 8 sources are
not anticipated to have maximum DMs much greater than this. This is
due to their location at relatively large Galactic latitudes away from
the Galactic plane (see Figure \ref{fig-2}), where there is generally
less Galactic plasma.  The DM spacing was slightly larger than the
ideal spacing of $\le$ 0.18 pc cm$^{-3}$ (see the single pulse search
described in Section 4.2), but it still maintained sensitivity to
sub-millisecond pulsars while making the Fourier search
computationally feasible.

Each dedispersed time series was Fourier transformed to produce a
power spectrum, and each spectrum was subsequently high-pass filtered
in order to remove any slowly varying noise contribution (red
noise). Each spectrum was also harmonically summed in order to enhance
signals with significant integer harmonic content (e.g., Taylor \&
Huguenin 1969\nocite{th69}). We performed a search on each
harmonically summed spectrum, with a S/N threshold of 7 and a
searchable modulation frequency range between 1600 Hz (100 Hz x 16
integer harmonics, corresponding to pulsars with fundamental periods
of 10 ms) and 10000 Hz. This restricted our full search sensitivity to
pulsars with $P < 10$ ms, which encompassed our period range of
interest. However, pulsars with periods greater than this may still
have been detectable if they had harmonics present above 100 Hz.

\begin{figure}
\centerline{\psfig{figure=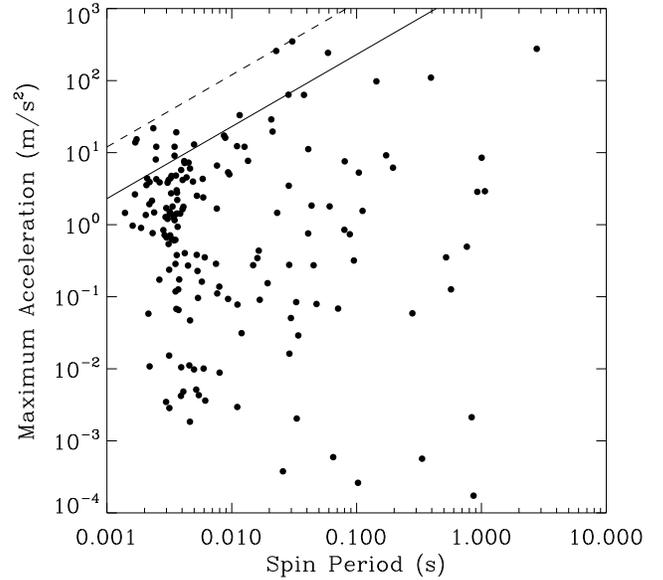,width=3.5in,angle=0}}
\caption{Maximum line-of-sight acceleration versus spin period for 167
known binary radio pulsars in the ATNF catalog with measured projected
semi-major axis values and binary orbital periods.  The maximum
acceleration was calculated using the method described by
\citet{fkl01}. The solid line represents the acceleration search range
$\pm$ 2.3 m s$^{-2}$($P$/1 {\rm ms}), for which full sensitivity was
maintained in our search. All but 13 of the pulsars fall below this
cutoff, indicating that our search range would be sufficient to detect
most known pulsars in binary systems even if they were observed while
undergoing maximum acceleration in their orbits. The dashed line
indicates the acceleration search range $\pm$ 12 m s$^{-2}$($P$/1 {\rm
ms}). This acceleration search range would provide full sensitivity to
all 167 pulsars shown in the plot. Our search was not performed out to
this range owing to computational limitations. However, given the
large flux densities of our sources compared to our sensitivity limit,
even such extremely accelerated pulsars would still be expected to be
detectable with a S/N above 7 in our search in the absence of flux
variability (see Section 4.1).\label{fig-3}}
\end{figure}

We conducted an acceleration search to account for any binary
acceleration, which causes a drift of the signal in phase over time
and a decrease in S/N.  Our search resulted in no sensitivity loss for
the acceleration range $\pm$ 2.3 m s$^{-2}$($P$/1 {\rm ms}), where $P$
is the candidate spin period.  For comparison, this search would have
ensured full sensitivity to most (154 out of 167) known binary radio
pulsars while they were undergoing maximum line-of-sight acceleration
in their orbits (see Figure \ref{fig-3}). This set of 167 pulsars was
taken from entries in the ATNF catalog which had measured values for
the projected semi-major axis and binary orbital period (see below),
but it did not include any binary systems recently discovered in the
High Time Resolution Universe (HTRU; Keith et al. 2010\nocite{kjv+10})
or Pulsar Arecibo L-Band Feed Array (PALFA; Cordes et
al. 2006\nocite{cfl+06}) surveys.

To calculate the maximum line-of-sight accelerations for these 167
pulsars, we used an expression derived by \citet{fkl01} in which the
acceleration of a pulsar in a binary system is described as a function
of observable parameters. From this equation, the maximum possible
acceleration $A_{max}$ is

\begin{equation}
A_{max} = \frac{4\pi^{2} \, a \, (1+e)^{2}}{P_{b}^{2} \, (1-e^{2})^2},
\end{equation}

where $a$ is the projected semi-major axis of the pulsar's orbit,
$P_{b}$ is the orbital period, and $e$ is the orbital eccentricity.

For the sample of 167 binary radio pulsars, if no eccentricity $e$ was
measured in the ATNF catalog, we assumed $e = 0$ (a perfectly circular
orbit) for the calculation.  As seen in Figure \ref{fig-3}, an
acceleration search range of $\pm$ 12 m s$^{-2}$ ($P$/1 {\rm ms})
would ensure full sensitivity to all known binary radio pulsars in the
sample undergoing maximum acceleration. However, the computational
time required to perform a search of our 75 sources with this full
acceleration range was excessive. Furthermore, the relatively large
flux densities of our sources ensured that pulsars with accelerations
near the highest values in this sample of 167 would probably still
have been detectable even if the S/N were reduced by acceleration bin
drift. The amount of Fourier bin drift of a pulse depends linearly on
the acceleration. For a acceleration of $\pm$ 12 m s$^{-2}$($P$/1 {\rm
ms}), the S/N would be reduced by a factor of 5.2 when searching only
out to an acceleration of $\pm$ 2.3 m s$^{-2}$($P$/1 {\rm ms}).  Since
we expect our weakest source to have a S/N of 44 in the absence of
time-dependent variability in intensity (see Section 3), the smallest
S/N we would expect from an extremely accelerated pulsar is $\ge$ 8,
which is still above our survey detection limit of 7 (see Figure
\ref{fig-1}).

The acceleration search produced a list of candidate signals for each
trial DM with periods and spectral S/N values. These candidate files
from different DM trials for a source were combined into a master
list. We subsequently removed candidates if they were deemed unlikely
to be pulsars (see below) and also removed duplicates and harmonically
related signals.  The heuristics used in keeping good candidates
included the requirements that the candidate have its maximum signal
at a DM greater than 1 pc cm$^{-3}$ and that it have a period $ P \ge$
0.5 ms.  The result was a list of the best pulsar candidates for a
source ranked by S/N. We then dedispersed and folded the raw data at
periods and DMs near the candidate values. We inspected the resulting
plots by eye to see if promising features were present, indicating
that a confirmation observation was warranted.

We tested this method with a 900 s observation of PSR B1937+21 using
the same observing system as was used for the 75 survey sources. We
successfully detected the pulsar in the search with a DM and period
consistent with its catalog values.

\subsection{Single Pulse Search}

We performed a search for single pulses in the data using SIGPROC.
Data from each source were dedispersed using 713 trial DMs with
variable spacing, with the DMs ranging between 0 and 100 pc
cm$^{-3}$. The maximum spacing between DM trials in this range was
0.18 pc cm$^{-3}$. We chose the spacing to ensure that the dispersion
smearing due to an offset in DM did not greatly exceed the dispersion
smearing within the frequency channels at that DM. We checked each
dedispersed time series for pulses having a range of widths using a
boxcar smoothing technique described by \cite{cm03}. We tested this
technique on a 900 s observation of the Crab pulsar (PSR B0531+21),
and we detected single pulses from the Crab easily. The method was
also tested on the same 900 s observation of PSR B1937+21 that was
used to test the periodicity search technique. No single pulses were
detected in that search, which had a minimum S/N threshold of 5. While
PSR B1937+21 and the Crab pulsar have similar giant pulse emission
rates \citep{cst+96} as well as comparable 1400 MHz flux densities,
the giant pulses emitted by PSR B1937+21 are much narrower than the
survey's sampling time of 61.44 $\mu$s (the pulse widths are on the
order of a few microseconds; see Figure 1 of Cognard et
al. 1996\nocite{cst+96}). These narrow pulses are smeared out within
each time sample, leading to a significant reduction in the S/N. Thus,
it is unlikely that such narrow pulses would be detected in our single
pulse search. This also indicates that our search is not sensitive to
extremely narrow single pulses or bursts (a few $\mu$s or less in
width) from our target sources.

\subsection{Follow-up Observations}

One candidate from the periodicity search showed merit and warranted
re-observation at the telescope. We conducted a confirmation
observation using the same observing system and integration time (900
s) as the original observation. The data were dedispersed and folded
at a range of DMs and periods near the candidate period and DM to see
if the same signal was re-detected. We also performed a blind
periodicity search and single pulse search on the confirmation
data. The signal was not confirmed.

\section{Results and Discussion}

We confirmed no dispersed pulsations or transient bursts from any of
the target sources in either the periodicity search or the single
pulse search.  Below we consider possible factors that could have led
to the non-detections if these sources were pulsars.

It is unlikely that diffractive scintillation caused us to miss
detections in most cases.  This can be shown using several different
assumptions to calculate the expected scintillation bandwidths for
these sources. First, an expression given by \citet{cwb85} gives an
estimated typical scintillation bandwidth $\Delta \nu_{scint}$ in MHz
(see also Equation 2 in Crawford et al. 2000\nocite{ckb00}):

\begin{equation}
\Delta \nu_{scint} \simeq  11 \, \nu_{c}^{4.4} \, d^{-2.2} .
\end{equation}

Here $\nu_{c}$ is the central observing frequency in GHz and $d$ is
the source distance in kpc. For $d$ = 0.2 kpc and $\nu_{c} = 1.2$ GHz,
$\Delta \nu_{scint}$ is comparable to our observing bandwidth of 800
MHz. Thus, sources with $d$ $<$ 0.2 kpc would be expected to
occasionally be undetectable in our search owing to diffractive
scintillation.  Note that this does not take into account the large
degree of variability in $\Delta \nu_{scint}$ that exists as a
function of sky position. A calculation of scintillation bandwidths
for the 75 sky positions of our sources using the NE2001 model
\citet{cl02} gives similar results: for a frequency of 1200 MHz and an
assumed distance $d$ = 0.2 kpc, we find that only 3 of the 75 sources
have $\Delta \nu_{scint} < 800$ MHz (where we have assumed a
Kolmogorov scaling dependence of $\nu^{4.4}$ for $\Delta
\nu_{scint}$). This number increases to 58 and 75 for $d = 0.5$ and
1.0 kpc, respectively. Thus, it appears that the range 0.2 to 0.5 kpc
represents a critical distance below which scintillation bandwidths
rapidly increase for these sources and where they can exceed the 800
MHz observing bandwidth. We can also estimate a typical source
distance using several assumptions. The median 1400 MHz flux density
of our 75 target sources is $S$ = 51 mJy, and the median value of the
1400 MHz pseudo-luminosity for known recycled radio pulsars in the
ATNF pulsar catalog \citep{mht+05} is $L$ $\sim$ 2 mJy kpc$^{2}$. The
pseudo-luminosity $L$ is defined according to $L$ = $S \, d^{2}$
\citep{lk04}.  Using these values for $L$ and $S$, we obtain a
characteristic source distance of $d$ $\sim$ 0.2 kpc, which is close
to the critical distance described above. The observed median $L$
value used here may be overestimated if the luminosity distribution of
the underlying population of recycled pulsars extends well below the
observed lower limit. In this case, the derived characteristic source
distance $d$ would be reduced as well.  In any case, even if $\Delta
\nu_{scint}$ were to exceed 800 MHz for every source, this would still
not prevent the detection of the majority of our sources. The
likelihood of detection of a source in this case is an exponential
function dependent on the source flux density and flux density limit
of the search observation. The total number of sources $n$ expected to
be detectable is the sum of these individual detection likelihoods
over all sources (see Crawford et al. (2000)\nocite{ckb00} for a
description and similar calculation),

\begin{equation}
n = \sum_{i=1}^{75} e^{-S_{i}/S_{min}} ,
\end{equation}

where $S_{min}$ is the 1200 MHz flux density limit of the survey
($\sim$ 3 mJy; see Figure \ref{fig-1}) and the sum is taken over all
75 sources having 1200 MHz flux densities $S_{i}$ (which we estimated
using the cataloged 1400 MHz values and a typical pulsar spectral
index of $\alpha$ = $-$1.6 in those cases where $\alpha$ could not be
measured). This calculation shows that $n = 71$ (out of 75) sources
should still be detectable in our survey. Therefore, diffractive
scintillation is not expected to be a dominant reason for the
non-detection of these sources if they are pulsars.

It is possible that some of these sources could be binary pulsars that
are being eclipsed and that the pulsar signal was masked at the time
of observation by material ablated from the companion by the pulsar
wind (e.g., black widows or redbacks).  As mentioned previously, very
rapidly spinning pulsars may exist more commonly in eclipsing systems
\citep{h08}. Since the obscuration is frequency-dependent, with
detectability decreasing at lower observing frequencies, this effect
could be mitigated by conducting higher-frequency observations in the
future. It is not likely that binary acceleration is a significant
factor in any of our non-detections. Our range of trial accelerations
in the periodicity search was sufficient to maintain sensitivity to
even the most highly accelerated radio pulsar systems currently known
(see Figure \ref{fig-3}).

Only 8 of our 75 sources had maximum estimated DMs from the NE2001
model that were greater than 100 pc cm$^{-3}$ \citep{cl02}. Since we
searched DMs up to 100 pc cm$^{-3}$, it is possible that dispersive
smearing rendered pulsations from some of these sources
undetectable. However, since the sources in our sample are very bright
relative to our survey detection limit, they are also likely to be
relatively close (see the estimated distance discussion above), with
corresponding DMs that are much smaller than the maximum DM. Thus,
dispersive smearing is unlikely to be a significant factor. Using the
known recycled pulsar population as a guide, the estimated
characteristic distance to the sources is small ($d$ $\sim$ 0.2 kpc;
see above). For this distance, the NE2001 model indicates that the
corresponding DMs are between about 1 and 5 pc cm$^{-3}$ for all 75
sources. Our candidate selection criteria in the search included
keeping Fourier candidates having DM $> 1$ pc cm$^{-3}$. Nevertheless,
pulsar signals at such low DMs are at risk of being flagged as RFI in
the processing. For several of our sources for which there was a
significant amount of RFI present in the data, removal of a low-DM
pulsar signal during the RFI excision process is a possibility. RFI in
these beams also affected the single pulse search, and any low-DM
transient signals from the target sources may have been removed along
with RFI.

If some of these sources are transient objects, such as RRATs
\citep{mll+06} or intermittent or nulling pulsars (e.g., Kramer et
al. 2006\nocite{klo+06}; Wang et al. 2007\nocite{wmj07}), it is
possible that they were simply not active at the time of
observation. The discovery of the intermittent emission behavior of
the supposedly ordinary pulsar PSR B1931+24 by \citet{klo+06} suggests
that many more such objects could have been missed in previous pulsar
searches. In this case, the chances of detecting pulsations from our
target sources would increase with additional observations, preferably
with longer observing times.

\section{Conclusions}

No new pulsars or transient objects were discovered in the survey, and
the question of what the unidentified target sources are remains
unanswered. Below we outline some speculative possibilities.

Some of these sources may be active galactic nuclei (AGN) that will be
identified in the future.  While only $\sim$ 10\% of known quasars and
BL Lac objects are more than 5\% linearly polarized \citep{ht99}, this
does not completely discount extragalactic sources as a
possibility. In fact, as seen in Table \ref{tbl-1}, 16 of the 92
previously unidentified sources searched by \citet{ckb00} have been
been identified as extragalactic objects and are no longer candidate
pulsars. The generally shallower spectral indices seen for our target
sources in Table \ref{tbl-1} relative to the steeper average observed
for the radio pulsar population also suggests this as a possibility.

Another possibility is that some of these sources might be
radio-emitting white dwarfs (WDs). Persistent radio emission has been
observed from several WD systems. \citet{cd82} were the first to
observe radio emission from AM Her, a magnetic cataclysmic variable
(MCV). AM Her is in a short-period binary orbit and has a strong
magnetic field. Radio emission was subsequently observed from two
other MCVs: AE Aqr \citep{bl87,bdc88} and AR UMa \citep{mg07}.
However, radio searches of a range of CV subclasses indicates that
they are not generally radio detected \citep{chm83,c87,mg07}.  It is
worth noting that \cite{mg07} observed the FIRST source J1023+0038 to
search for radio emission. At the time, this source was tentatively
identified as a CV associated with a FIRST radio source
\citep{bwb+02}. \cite{mg07} found no radio emission from J1023+0038,
but the source was subsequently detected and identified as the first
redback radio pulsar system discovered in the Galactic field
\citep{asr+09}.  This highlights the importance of repeated
observations of undetected radio sources.

The observed radio sources might also be radio pulsars having closely
aligned spin and magnetic axes.  A study by \citet{tm98} found that
there is a tendency for the magnetic and spin axes to become aligned
as pulsars age. Using pulsar polarization data sets from \citet{r93}
and \cite{g94}, they found that this alignment occurs on a time-scale
of $\sim 10^{7}$ yr. More recent work by \cite{ycb+10} using
Candy-Blair evolution models \citep{cb83,cb86,j76} also shows
progressive alignment occurring, but on a time-scale of $\sim 10^{6}$
yr. Such aligned rotators would be hard or impossible to detect in
traditional pulsar searches owing to their decreased (or non-existent)
modulation \citep{tm98}. This is illustrated in Figure 1 of
\citet{ycb+10}, which shows that in cases where the emission cone
encompasses the spin axis, one would not expect to see a modulated
signal. If our target sources are indeed unmodulated radio pulsars
with aligned magnetic and spin axes, then these models indicate that
they are likely to be old pulsars. In this case, the shallower
spectral indices of these older pulsars (see Table \ref{tbl-1})
relative to the general radio pulsar population would need to be
explained.

If these unidentified sources are rapidly rotating pulsars with
modulated radio emission, the effects of diffractive scintillation,
dispersive smearing, and binary acceleration are unlikely to have
prevented the detection of a large majority of the targets in our
search. However, we cannot rule out eclipsing, intermittent emission,
nulling, or RFI as possibly important factors. Just as {\it Fermi}
sources must be searched multiple times before declaring that no radio
pulsar is present \citep{rap+12}, repeated observations of our targets
may be required to reach the same conclusion.

\acknowledgments

The National Radio Astronomy Observatory (NRAO) is a facility of the
National Science Foundation operated under cooperative agreement by
Associated Universities, Inc.  Funding for some of the equipment used
was provided by West Virginia EPSCoR and Research Corporation.
D.R.S. was supported at the NRAO by the Research Experience for
Undergraduates program, which is funded by the National Science
Foundation. This research has made use of the SIMBAD database,
operated at CDS, Strasbourg, France. We thank Maura McLaughlin for
valuable suggestions which helped to optimize the periodicity search
algorithm, and we thank an anonymous referee for recommendations that improved
the paper. D.R.S. also wishes to thank the staff of the Green Bank
Observatory for their hospitality during her time there.

\LongTables

\begin{deluxetable}{lccccccc}

\renewcommand{\arraystretch}{0.75}
\tablecaption{NVSS Radio Source Targets\label{tbl-1}}
\tablewidth{0pt}
\tablehead{
\colhead{NVSS Source} &
\colhead{$\alpha$ (J2000.0)} &
\colhead{$\delta$ (J2000.0)} &
\colhead{$S_{1400}$\tablenotemark{a}} &
\colhead{Lin. Poln\tablenotemark{b}} &
\colhead{Source ID\tablenotemark{c}} &
\colhead{$\alpha$\tablenotemark{d}} &
\colhead{Survey\tablenotemark{e}} \\
\colhead{} &
\colhead{(h m s)} &
\colhead{(d m s)} &
\colhead{(mJy)}  &
\colhead{(\%)}  &
\colhead{} &
\colhead{} &
\colhead{}}

\startdata
J000240$-$195252 &  00 02 40.96 &  $-$19 52 52.3  & 60  & 9   &           &           &                  \\	   
J000404$-$114858 &  00 04 04.90 &  $-$11 48 58.4  & 459 & 6   &   B       & 	      & 	         \\
J001109$-$225458 &  00 11 09.91 &  $-$22 54 58.5  & 38  & 10  &      	  & 	      & 	         \\
J001444$-$280047 &  00 14 44.06 &  $-$28 00 47.3  & 54  & 14  &      	  & 	      & 	         \\
J002330$-$215537 &  00 23 30.21 &  $-$21 55 37.6  & 137 & 8   &           & $-$0.32(11) &    PMN           \\
J002449$+$030834 &  00 24 49.37 &  $+$03 08 34.7  & 69  & 9   &      	  & 	      & 	         \\
J002651$-$111252 &  00 26 51.45 &  $-$11 12 52.4  & 169 & 8   &   Q       & 	      & 	         \\
J002702$-$303032 &  00 27 02.07 &  $-$30 30 32.0  & 24  & 14  &      	  & 	      & 	         \\
J003233$-$264917 &  00 32 33.03 &  $-$26 49 17.6  & 135 & 7   &   Q       & 	      & 	         \\
J003708$-$232340 &  00 37 08.81 &  $-$23 23 40.5  & 67  & 6   &   	  & 	      & 	         \\
J004021$+$132937 &  00 40 21.80 &  $+$13 29 37.9  & 34  & 10  &   	  & 	      & 	         \\
J005151$+$022944 &  00 51 51.30 &  $+$02 29 44.2  & 15  & 21  &   	  & 	      & 	         \\
J005410$-$175413 &  00 54 10.78 &  $-$17 54 13.0  & 30  & 11  &   	  & 	      & 	         \\
J005736$+$134145 &  00 57 36.44 &  $+$13 41 45.4  & 65  & 9   &   G       & 	      & 	         \\
J010711$-$121123 &  01 07 11.79 &  $-$12 11 23.6  & 60  & 6   &   	  & 	      & 	         \\
J011448$-$321951 &  01 14 48.89 &  $-$32 19 51.7  & 124 & 16  &           & $-$0.37(13) &    PMN           \\
J013840$-$295445 &  01 38 40.50 &  $-$29 54 45.9  & 46  & 10  &   	  & 	      & 	         \\
J014614$+$022208 &  01 46 14.62 &  $+$02 22 08.2  & 138 & 8   &           & $-$0.13(9)  &    PMN           \\	   
J014727$+$071502 &  01 47 27.77 &  $+$07 15 02.9  & 241 & 6   &           & $-$0.72(6)  &    TXS           \\	   
J015456$-$242233 &  01 54 56.90 &  $-$24 22 33.5  & 45  & 11  &      	  & 	      & 	         \\
J021459$+$102748 &  02 14 59.23 &  $+$10 27 48.8  & 29  & 13  &      	  & 	      & 	         \\
J021750$-$235456 &  02 17 50.76 &  $-$23 54 56.3  & 90  & 11  &      	  & 	      & 	         \\
J022333$+$073219 &  02 23 33.94 &  $+$07 32 19.6  & 125 & 13  &   G       & 	      & 	         \\
J022340$+$115910 &  02 23 40.83 &  $+$11 59 10.2  & 34  & 10  &      	  & 	      & 	         \\
J022441$+$135733 &  02 24 41.85 &  $+$13 57 33.1  & 95  & 10  &           & $-$0.17(13) &    87GB          \\
J023855$-$303202 &  02 38 55.19 &  $-$30 32 02.6  & 165 & 5   &    	  & 	      & 	         \\
J024944$+$123706 &  02 49 44.50 &  $+$12 37 06.3  & 261 & 6   &           & $-$0.65(12) &    87GB          \\
J025106$-$174239 &  02 51 06.22 &  $-$17 42 39.5  & 69  & 13  &           & $-$0.09(15) &    PMN           \\
J025805$-$314627 &  02 58 05.95 &  $-$31 46 27.8  & 247 & 8   &   Q       & 	      & 	         \\
J025927$+$074739 &  02 59 27.06 &  $+$07 47 39.2  & 834 & 5   &   Q       & 	      & 	         \\
J025904$+$470840 &  02 59 04.20 &  $+$47 08 40.4  & 108 & 9   &           & $-$0.58(7)  &    TXS           \\	   
J031726$+$060614 &  03 17 26.85 &  $+$06 06 14.7  & 200 & 8   &           & $-$0.89(14) &    PMN           \\
J032213$-$345833 &  03 22 13.10 &  $-$34 58 33.2  & 49  & 7   &      	  & 	      & 	         \\
J032615$-$324324 &  03 26 15.12 &  $-$32 43 24.3  & 95  & 5   &      	  & 	      & 	         \\
J034914$+$035445 &  03 49 14.31 &  $+$03 54 45.4  & 150 & 8   &           & $+$0.30(16) &    PMN           \\
J040342$+$644556 &  04 03 42.80 &  $+$64 45 56.1  & 74  & 5   &           & $-$0.22(10) &    87GB          \\
J042119$+$351115 &  04 21 19.72 &  $+$35 11 15.6  & 68  & 10  &      	  & 	      & 	         \\
J045828$+$495355 &  04 58 28.75 &  $+$49 53 55.8  & 19  & 13  &      	  & 	      & 	         \\
J050554$+$260625 &  05 05 54.20 &  $+$26 06 25.1  & 23  & 11  &      	  & 	      & 	         \\
J051843$+$643958 &  05 18 43.68 &  $+$64 39 58.0  & 28  & 13  &      	  & 	      & 	         \\
J060650$+$440140 &  06 06 50.20 &  $+$44 01 40.9  & 147 & 8   &           & $-$0.24(5)  &    7C            \\	   
J060718$+$291527 &  06 07 18.95 &  $+$29 15 27.7  & 25  & 16  &      	  & 	      & 	         \\
J062052$+$733441 &  06 20 52.10 &  $+$73 34 41.2  & 85  & 10  &           & $-$0.69(11) &    87GB          \\
J070120$+$263157 &  07 01 20.74 &  $+$26 31 57.1  & 32  & 12  &      	  & 	      & 	         \\
J071923$+$293551 &  07 19 23.05 &  $+$29 35 51.8  & 18  & 13  &      	  & 	      & 	         \\
J073313$+$333151 &  07 33 13.31 &  $+$33 31 51.8  & 19  & 14  &      	  & 	      & 	         \\
J075501$+$301347 &  07 55 01.81 &  $+$30 13 47.4  & 51  & 12  &           & $-$1.13(9)  &    TXS           \\	   
J075536$+$334159 &  07 55 36.71 &  $+$33 41 59.2  & 82  & 7   &      	  & 	      & 	         \\
J075752$+$272111 &  07 57 52.82 &  $+$27 21 11.3  & 45  & 7   &      	  & 	      & 	         \\
J075808$+$392928 &  07 58 08.84 &  $+$39 29 28.7  & 545 & 8   &   G       & 	      & 	         \\
J080212$+$312240 &  08 02 12.78 &  $+$31 22 40.7  & 86  & 10  &      	  & 	      & 	         \\
J080519$+$273736 &  08 05 19.02 &  $+$27 37 36.1  & 42  & 10  &      	  & 	      & 	         \\
J080601$+$331010 &  08 06 01.70 &  $+$33 10 10.3  & 44  & 6   &      	  & 	      & 	         \\
J081040$+$303433 &  08 10 40.30 &  $+$30 34 33.7  & 154 & 6   &      	  & 	      & 	         \\
J084030$+$292336 &  08 40 30.72 &  $+$29 23 36.8  & 18  & 14  &     	  & 	      & 	         \\
J084308$+$373816 &  08 43 08.76 &  $+$37 38 16.2  & 111 & 12  &     	  & 	      & 	         \\
J084456$+$362927 &  08 44 56.27 &  $+$36 29 27.5  & 50  & 7   &    	  & 	      & 	         \\
J084647$+$374615\tablenotemark{f} &  08 46 47.43 &  $+$37 46 15.1  & 22  & 16  &   	  & 	      & 	         \\
J090305$+$352316 &  09 03 05.50 &  $+$35 23 16.2  & 58  & 6   &   	  & 	      & 	         \\
J091147$+$334917 &  09 11 47.77 &  $+$33 49 17.1  & 380 & 7   &   B       & 	      & 	         \\
J092329$+$301106 &  09 23 29.97 &  $+$30 11 06.7  & 36  & 6   &   Q       &           & 	         \\
J092822$+$414221 &  09 28 22.18 &  $+$41 42 21.9  & 98  & 12  &           & $-$0.95(15) &    TXS           \\
J094459$+$380317 &  09 44 59.21 &  $+$38 03 17.6  & 43  & 12  &      	  & 	      & 	         \\
J100022$+$371844 &  10 00 22.29 &  $+$37 18 44.4  & 35  & 10  &      	  & 	      & 	         \\
J100357$+$324403 &  10 03 57.63 &  $+$32 44 03.7  & 429 & 8   &   Q       & 	      & 	         \\
J101349$+$344550 &  10 13 49.59 &  $+$34 45 50.7  & 356 & 6   &   Q       & 	      & 	         \\
J103319$+$285121 &  10 33 19.66 &  $+$28 51 21.0  & 29  & 10  &           &  0.00(17) &    87GB          \\
J112612$+$341818 &  11 26 12.32 &  $+$34 18 18.2  & 40  & 7   &      	  & 	      & 	         \\
J112951$+$362217 &  11 29 51.56 &  $+$36 22 17.1  & 122 & 7   &      	  & 	      & 	         \\
J114523$+$314515 &  11 45 23.20 &  $+$31 45 15.3  & 83  & 10  &      	  & 	      & 	         \\
J114608$+$260105 &  11 46 08.52 &  $+$26 01 05.3  & 116 & 7   &           & $-$0.62(14) &    87GB          \\
J115043$+$302018 &  11 50 43.89 &  $+$30 20 18.3  & 32  & 14  &      	  & 	      & 	         \\
J120125$+$255006 &  12 01 25.63 &  $+$25 50 06.9  & 25  & 26  &      	  & 	      & 	         \\
J120144$+$312904 &  12 01 44.50 &  $+$31 29 04.1  & 91  & 6   &      	  & 	      & 	         \\
J122004$+$311149 &  12 20 04.37 &  $+$31 11 49.0  & 29  & 10  &   Q       & 	      & 	         \\
J123454$+$291744 &  12 34 54.36 &  $+$29 17 44.2  & 446 & 9   &           & $-$0.52(9)  &    7C            \\	   
J123650$+$370603 &  12 36 50.94 &  $+$37 06 03.8  & 63  & 10  &      	  & 	      & 	         \\
J124219$+$272156 &  12 42 19.75 &  $+$27 21 56.2  & 70  & 10  &           & $-$0.65(8)  &    7C            \\
J125124$+$364354 &  12 51 24.28 &  $+$36 43 54.9  & 30  & 8   &      	  & 	      & 	         \\
J133426$+$343425 &  13 34 26.94 &  $+$34 34 25.8  & 49  & 8   &      	  & 	      & 	         \\
J134324$+$290358 &  13 43 24.38 &  $+$29 03 58.6  & 24  & 18  &      	  & 	      & 	         \\
J141440$+$402229 &  14 14 40.46 &  $+$40 22 29.7  & 44  & 6   &      	  & 	      & 	         \\
J142658$+$403539 &  14 26 58.42 &  $+$40 35 39.6  & 29  & 9   &   Q       & 	      & 	         \\
J143447$+$380514 &  14 34 47.05 &  $+$38 05 14.4  & 153 & 9   &   A	  & 	      & 	         \\
J145844$+$372022 &  14 58 44.77 &  $+$37 20 22.0  & 215 & 6   &   B       & 	      & 	         \\
J150808$+$281811 &  15 08 08.32 &  $+$28 18 11.2  & 77  & 7   &      	  & 	      & 	         \\
J154740$+$395438 &  15 47 40.19 &  $+$39 54 38.4  & 131 & 15  &      	  & 	      & 	         \\
J160616$+$270930 &  16 06 16.20 &  $+$27 09 30.6  & 29  & 8   &      	  & 	      & 	         \\
J160950$+$262839 &  16 09 50.12 &  $+$26 28 39.4  & 20  & 17  &      	  & 	      & 	         \\
J161827$+$293118 &  16 18 27.05 &  $+$29 31 18.3  & 30  & 9   &      	  & 	      & 	         \\
J163552$+$375159 &  16 35 52.59 &  $+$37 51 59.5  & 48  & 7   &      	  & 	      & 	         \\
J232102$-$175822 &  23 21 02.41 &  $-$17 58 22.0  & 18  & 15  &     	  &           &                  \\
\enddata

\tablecomments{The original set of 92 NVSS target sources from Table 1
of \citet{ckb00} is presented here, including 16 sources that have
been more recently identified as extragalactic objects.  The flux
densities and polarization fractions listed were taken from the NVSS
catalog browser (revision 2.18, 2004-02-03;
http://www.cv.nrao.edu/nvss/NVSSlist.shtml).}

\tablenotetext{a}{Nominal 1400 MHz flux density from the NVSS
catalog.}

\tablenotetext{b}{Linearly polarized intensity as a percentage of
total source intensity, derived from the NVSS catalog.}

\tablenotetext{c}{Source identifications from SIMBAD. The
identification codes are: A = AGN; B = BL Lac object; G = Galaxy; Q =
Quasar.}

\tablenotetext{d}{Spectral index $\alpha$ defined according to $S \sim
\nu^{\alpha}$. The figure in parentheses represents the uncertainty in
the last digit quoted in the measured value.}

\tablenotetext{e}{Radio survey used with the NVSS to determine the
spectral index, where possible.  The survey codes are: 7C = Seventh
Cambridge Catalog at 151 MHz \citep{hrw+07}; TXS = The Texas Survey at
365 MHz \citep{dbb+96}; 87GB = 87 Green Bank Catalog at 4850 MHz
\citep{gc91}; PMN = Parkes-MIT-NRAO Survey at 4850 MHz \citep{gw93}.}

\tablenotetext{f}{Source not observed owing to telescope
scheduling limitations.}

\end{deluxetable}

\begin{deluxetable}{lccccc}
\renewcommand{\arraystretch}{1.00}
\tablecaption{Observing Parameters for Jodrell Bank and NRAO 43-m Surveys  \label{tbl-2}}
\tablewidth{0pt}
\tablehead{
\colhead{Survey} &
\colhead{\citet{ckb00}} & 
\colhead{This Survey} 
\newline
}
\startdata

Number of Targets                          & 92                    & 75    \\
Telescope                                  & Lovell (Jodrell Bank) & NRAO  \\
Telescope diameter (m)                     & 76                    & 43    \\
Central observing frequency, $\nu_{c}$ (MHz) & 610                 & 1200  \\
Observing bandwidth (MHz)                  & 1                     & 800   \\
Number of frequency channels               & 32                    & 4096  \\
Number of bits per sample                  & 1                     & 8     \\ 
Integration time per source (s)            & 420                   & 900   \\
Sampling time ($\mu$s)                     & 50                    & 61.44 \\
\enddata

\end{deluxetable}

\end{document}